\documentclass[useAMS,usenatbib,usegraphicx]{mn2e}

\newcommand{\mum}{$\mu$m}

\title[I. Collisional evolution and reddening of asteroid
  surfaces]{I. Collisional evolution and reddening of asteroid surfaces:
  The problem of conflicting timescales and the role of
  size-dependent effects} 
\author[S. Marchi et al.]{S. Marchi$^{1}$, P. Paolicchi$^{2}$ and D.~C. Richardson$^{3}$\\ 
$^{1}$Departement Cassiop\'{e}e,  Universite de Nice  - Sophia
  Antipolis, Observatoire  de la C\^{o}te d'Azur,  CNRS, Nice, France\\
$^{2}$Dipartimento di Fisica, Universit\`a di Pisa, 56127 Pisa, Italy\\ 
$^{3}$Department of  Astronomy, University of Maryland, College Park, MD 20742 USA}
\begin{document}

\date{Accepted . Received ; in original form }

\pagerange{\pageref{firstpage}--\pageref{lastpage}} \pubyear{}

\maketitle

\label{firstpage}

\begin{abstract}

Space weathering is the generic term used for processes that modify
the optical properties of surfaces of atmosphereless rocky bodies
under exposure to the space environment.  The general agreement about
the relevance of the effects of space weathering on the spectral
properties of S-complex asteroids fails when some basic quantitative
estimates are attempted.  In particular, there is severe disagreement
regarding the typical timescales for significant spectral reddening to
occur, ranging from 1~Myr to 1~Gyr.\\
\noindent
Generally speaking, the spectral reddening of an individual object can
be considered as the sum of three terms, one (which is relevant for
statistical analyses) depending on the exposure of the object to space
weathering during its lifetime, a second one due to the original
surface composition, and a third one (a {\it noise} term) due to the
combination of poorly constrained effects (e.g., structure and texture
of the surface). \\
\noindent
The surface of an asteroid is usually covered by regolith, and its
presence and properties presumably play a critical role in the
weathering processes. In this paper we discuss the role played by
collisional evolution in affecting the spectral properties of
asteroids and refreshing the surfaces due to the formation of ejecta,
and the necessity of a simultaneous modeling of collisions and
weathering processes.  We introduce a new idea, based on the
possibility of a sort of saturation of the refreshing process whenever
a massive reaccumulation of the impact ejecta takes place. In this
case, a dependence of the overall reddening on the asteroid size
should naturally come out. We show that this conclusion is indeed
supported by available main belt asteroid spectroscopic data.\\

\end{abstract}

\begin{keywords}
minor planets, asteroids: general; meteorites, meteors, meteoroids
\end{keywords}

\section{Introduction}
\indent

The spectral reddening of asteroidal surfaces is of fundamental
importance to understanding the spectral properties derived from
remote observations of asteroids and for making a reliable comparison
with the corresponding laboratory meteorite data. The process of
spectral reddening for S-complex asteroids has been recognized for a
long time \citep{cha73,gaf93}, and it has been attributed to the
alteration of surface optical properties under exposure to the space
environment.  More recently, several contributions from laboratory
experiments \citep{hir01,str05}, in addition to dedicated asteroid
observations and modeling, have disclosed several intriguing aspects
of the reddening processes \citep{jed04,mar06a,pao07}. It has also been
shown that, to some extent, the overall results obtained from
S-complex asteroids may play a non-negligible role also for other main
taxonomic complexes and spectral types \citep{laz06,mar10}.\\
\noindent
On the basis of a statistical analysis of a large sample of visible
spectra of S-complex asteroids \cite{mar06a} (hereinafter P1),
suggested that the solar wind is the dominant cause of the
reddening. This conclusion was based on the significant dependence of
the spectral slope on the exposure ($E$) to the solar wind, namely the
integrated ion flux that an asteroid received from the Sun during its
past evolution. This analysis is based on a large sample of both
near-Earth asteroids (NEAs) and main belt asteroids (MBAs). In a
series of papers \citep{mar06b,pao07}, additional details of the
process were determined, and an estimate of the reddening timescale
was given. It was found that 80\% of the slope reddening is reached
after about 200~Myr at 1~AU; or 800~Myr at 2~AU \citep{pao07}.  In
general, the derived correlation between the slope and the asteroid
age, the latter estimated by the collisional lifetime, requires that
the time derivative exhibit an exponential decay: a fast initial
reddening (leading to almost 50\% of the final reddening in a few tens
Myr) is followed by a slower increase, asymptotically pointing to
saturation. Note, however, that the estimated reddening timescale due
to heavy ion bombardment is of the order of $0.01-1$~Myr at 1~AU
\citep{str05}.\\
\noindent
A recent analysis based on S-complex MBA families \citep[][hereinafter
  P2]{ver09} claims that the solar wind is the dominant cause of the
reddening.  This conclusion was derived on the basis of the fast
reddening timescale ($\sim 1$~Myr) observed for two young asteroid
families.  Moreover, a slower residual reddening is present in their
data for older families. Despite P2 agrees on the what already found
by P1 about the solar wind-driven space weathering, the derived
timescales differs by orders of magnitude.  \\
\noindent
To make the story more complicated, a recent analysis \citep{wil10}
suggests that space weathering timescales may be very large (of
the order of $10^2-10^3$~Myr for MBAs), claiming agreement with a
different series of laboratory experiments that assume 
micrometeorite bombardment is the dominant weathering process
\citep[][and references therein]{wil10}.  While the timescales found by
Willman et al.\ are partially in agreement with some P1 results, the
physical process that is the claimed cause of the reddening is
completely different.\\
\noindent
A recent comprehensive review by \cite{gaf10} points out that the
consequences of space weathering may vary strongly among bodies, due,
possibly, both to different physical processes and to different
properties of the affected surfaces. For instance, it has been found
that the space weathering of lunar regolith is likely due to the
formation of nanophase iron particles \citep{pie00} produced by the
vaporization of surface particles by microimpacts.  It is noteworthy
to recall that lunar ray craters (e.g. Tycho and Copernicus) show a
moderate reddening even if their estimated age (or the corresponding
exposure to the Sun) is relatively large, of the order of hundreds of
Myr \citep{sto01}.  Microimpact vaporization, however, might have
nothing to do with the reddening of asteroids, perhaps due to much
lower mean impact speeds in the main belt (5~km/s) compared to that on
the Moon (18~km/s).  Moreover, even if we decide to restrict our
analysis to the asteroids, severe differences come out in the
weathering properties of objects that should be -in principle- rather
similar (e.g., Eros and Ida).  According to conclusions of
\cite{gaf10}, it is not easy, and probably potentially misleading, to
define a simple -uniparametric- asteroid weathering scenario. \\
\noindent
On the other hand, both the solar wind and micrometeorite bombardment
weathering processes coexist for all asteroids, even if their relative
efficiency may vary depending on the heliocentric distance.  The
different effects of space weathering should be foremost dependent on
the different properties of the impacted surfaces (composition,
texture, presence of regolith and so on). Unless we are able to find a
well-defined selection criterion, which divides the asteroids into
different groups, with different values of all the variables affecting
the individual reaction of the bodies to the space environment, the
weathering properties of {\it individual} bodies are difficult to
assess, but the possibility of a statistical analysis and the search
for a unitary statistical model remains open.\\
\noindent
In this series of papers we discuss the role of mini- and
micro-collisions, as a potential solution to the two-timescales
conundrum \citep[as already suggested, at a very qualitative level,
  in][]{pao09}.\\
\noindent
The role of gardening in affecting weathering properties has been
discussed in the literature \citep[see, for instance,][]{gil02}, and a
quantitative model has been introduced by \cite{wil08,wil10}.
However, their models, based on some ad hoc assumptions (see later)
seem to not be capable of fitting together the properties of the
asteroids with those of ordinary chondrites (OCs). In a recent improvement
\citep{wil11}, the fit with ordinary chondrites could be attained, but
the reddened surfaces of young asteroid families do not fit with the
inferred space weathering timescale.
\noindent
Note that, however, not all S-type asteroids are expected to be
genetically linked to ordinary chondrites. Indeed, several observed
S-types have inferred compositions incompatible with ordinary
chondrites \citep{gaf93,mar05}; while laboratory analysis show that
OCs may have originated from a comparatively small number of parent
bodies \citep{bur02}. Therefore, we caution that the average
unweathered spectral slope of S-types may be different from that of
ordinary chondrites.\\
\noindent
Apart from the technical details, the essence of the problem can be
qualitatively sketched in the following way. We have, on one side,
some evidence for a short ($\sim1$~Myr) timescale for substantial
reddening: experiments on ion implantation and the observed reddening
of some of the youngest asteroid family members.  On the other side,
we have also hints for a longer timescale ($\sim10^2-10^3$~Myr), both
from micrometeorite bombardment experiments and from a significant
continuation of the reddening on old asteroids.  However, at a
microphysical scale, both processes must be present simultaneously.
Thus if the short timescale holds, there is no conflict, while, in the
opposite case, one should understand why the faster process does not
work. Moreover, if we introduce a de-weathering effect due to
collisions, we can have two extreme cases: if the collisional
timescale is by far shorter than that of the weathering, no
significant reddening can happen at all; if it is by far longer, the
collisions do not much affect the reddening.  Thus, the role of
collisions, within this scheme, may be relevant only if the gardening
has a timescale comparable to the weathering; this may be the case, as
suggested by \cite{wil10}, but this is not enough to solve the
two-timescales conundrum. \\
\noindent
Thus, the solution may require an additional effect to be
considered. In our model, this effect is connected to the {\it
  saturation} of the spectral properties, which can take place even in
the presence of frequent gardening of the surface, due to the
reaccumulation of the ejecta created by an impact; this reaccumulation
can partially cover the surface with already reddened layers. The
relevance of the reaccumulation may differ in importance, depending on
the target's size: thus a different behavior for small vs.\ large
objects has to be expected. \\
\noindent
In this paper, we show new observational results in support of the
above mechanism. Detailed numerical simulations of the regolith
evolution and impact processes will be deferred to the next paper in
the series.\\

\section{Conflicting timescales}

\subsection{The ``short-time'' camp and the problem of near-Earth asteroids}
\indent

The most explicit statement of a fast reddening of S-complex asteroids
is presented in P2, where the analysis is restricted to a set of MBAs,
namely members of dynamical families.  The reddening timescale is
obtained, in P2, by the correlation of the spectral slope of asteroids
(corrected to eliminate compositional effects) and their family age.
The timescale obtained, for most of the reddening, using data
concerning two young MBA families (namely, Datura and Lucascavin), is
of the order of 1~Myr. This value is consistent with the estimates
from ion bombardment laboratory experiments \citep{str05}, therefore
the authors of P2 claim that the reddening of young asteroid families
is due to the solar wind. However, they find also a slower further
increase of the spectral slope, which, according to the authors, might
be due to different physical processes.  Note that the observed
initial fast reddening is based on the assumption that the unweathered
slopes of the S-type Datura and Lucascavin families are the same as
that observed in ordinary chondrites. Otherwise, the resulting
reddening time dependence may be different, even showing no intrinsic
evidence of a double timescale. However, the existence of a fast
reddening process and the observed peculiar properties of Q-type NEAs
\citep{mar06b,bin10} do not support this extreme possibility.\\
\noindent
The reddening, if mainly caused by the Sun, has to be
related to the exposure $E$ to the solar wind, as defined in
\cite{mar06a}:
$$ E \simeq \frac{1}{a^2 (1-e^2)^{1/2}} \cdot {\rm age}. $$

\begin{figure}
\vspace{0.5cm}
\includegraphics[width=9cm,angle=-90]{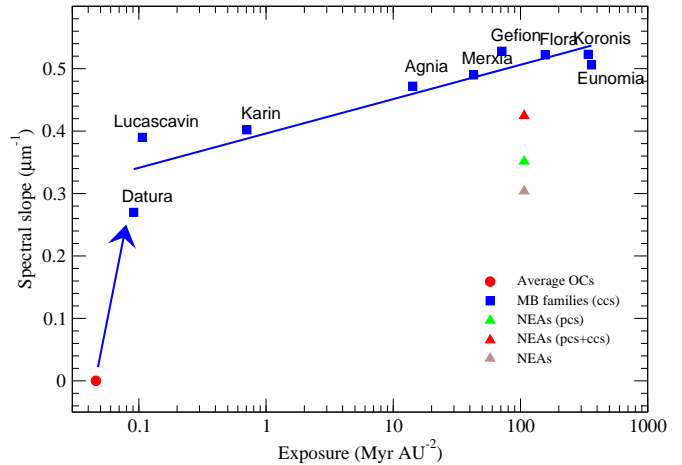}
\caption{A slope-exposure relation corrected for chemical composition,
  based on the data of Vernazza et~al.~(2009). Average NEA datapoints
  are also included (triangles), represented with its nominal age as
  estimated in Marchi et al.~(2006a, 2006b) and Paolicchi et
  al.~(2007).  The three points correspond to the nominal mean slope,
  to the same data corrected for the perihelion term (labeled as
  ``pcs''), and also with a further compositional correction due to
  the different mean value of the olivine-pyroxene parameter between
  MBAs and NEAs (labeled as ``pcs+ccs''). See text for further
  details.}
\label{f1}
\end{figure}

For the data sample used in P2, the slope-age and slope-$E$ plots are
rather similar, since the range of heliocentric distances is rather
narrow ($2.23-2.87$~AU). The slope-$E$ relation, obtained from the
same data used by \cite{ver09}, is represented in Fig.~\ref{f1}. On
the same plot, we also report the average data point from the NEAs
dataset of \cite{pao07}.  For a more detailed comparison between NEAs
and MBAs, we correct the average NEA's data point for surface
composition and perihelion de-reddening effects due to close
encounters with terrestrial planets.  For the composition effect, a
mean relative abundance of olivine and pyroxene $ol/(ol+opx)\sim0.7$
was estimated for NEAs \citep{ver08}. Thus, using the formula given in
\cite{ver09} to correct the slope for a different composition, the
mean spectral slope of NEAs has to be slightly increased by
$+0.07$\mum$^{-1}$ (Fig.~\ref{f1}). This data point was further
corrected in order to take into account the perihelion de-reddening
correction, as introduced in \cite{pao07}.\\
\noindent
We see that, in all cases, the NEA data points are too low compared to
MBAs for a similar exposure. On the other hand, the average NEA slope
is similar to those of the Datura and Lucascavin families, thus
implying a similar exposure. Due to the different heliocentric
distance, this exposure corresponds to a typical age of the order of
0.1~Myr for NEAs\footnote{A similar conclusion is obtained by also
  rescaling the difference in exposure between the young MBA families
  and a typical NEA ($a \sim 1.5$~AU; $e \sim 0.4$).  The reddening
  timescale of 1~Myr suggested by P2 becomes $0.1-0.2$~Myr for a
  typical NEA.}.\\
\noindent
 This timescale is very short compared to all the relevant
evolutionary timescales concerning NEAs:
\begin{enumerate}
\item{their average ages (or collisional lifetimes) taking also into
  account the time passed in the main belt \citep[the latter may
      exceed the dynamical lifetime of a typical NEA and have a
      dominant role in determining the reddeding; see][]{mar06a};}
\item{their typical lifetimes as NEAs;}
\item{the time interval between significant de-weathering close
  encounters \citep{mar06b,bin10}.}
\end{enumerate}
In other words, the NEAs should be almost completely reddened, a
conclusion that is in stark contradiction with observations.\\
\noindent
A drastic solution to the problem assumes that the short timescale for
reddening estimated by P2 is a fluke, and the correct timescale is of
the order of hundreds of Myr. In this case, the NEAs problem would be
solved; however, to match ordinary chondrites and young family objects
is far less easy. In fact, the typical slope of the youngest known
asteroids is, for the most part, significantly redder than that of the
ordinary chondrites. \\
\noindent
It is also possible to suggest that the timescale estimated in P2, and
based only on a couple of observational points, is underestimated by,
say, one order of magnitude. In this case the de-weathering effect due
to close planetary passes \citep{mar06b,bin10} might do the job.  This
possibility should be enforced by weakening, as discussed before, the
link between the slopes of ordinary chondrites and unweathered
S-types.  \\
\noindent
Another, and more intriguing, possibility, will be discussed in
greater detail below, namely that the main difference between NEAs and
MBAs should be the {\bf size}: most NEAs are small and, if the
gardening refreshing is rather fast but can be significantly weakened
by the self-reaccumulation of the ejecta, then small bodies cannot
become too red. See Sect.~3 for further discussion.\\

\subsection{The ``long-time'' camp}

The long-time camp is mainly represented by \cite{wil08,wil10}.  Their
timescales are of the order of several hundreds of Myr, and a
different laboratory counterpart (namely, micrometeorite impacts) is
suggested \citep{hir01}. The long-time camp is partially supported
also by the results of our group (P1). We have found that a residual
reddening (20\%) takes place over very long ages, even of the order of
1~Gyr. However, in our model most of the reddening takes place in the
first period, according to the analysis and plots presented in
\cite{pao07}.\\
\noindent
Moreover, there is evidence for a fast reddening of some young objects
(such as those of the Karin family), in substantial agreement with the
short-time camp suggestions \citep{pao07}. Finally, the apparent
Sun-dependence of the weathering (the slope is more strongly dependent
on the exposure than on the age: \cite{mar06a}) strongly supports the
dominance of the solar wind-driven ion-implantation processes, even if
it cannot be considered as final, unequivocal proof: in fact, the
micrometeoritic bombardment may also depend on the distance from the
Sun \citep{cin92}, as claimed by \cite{wil10}. Note, also, that the
total fluxes used in experiments are tuned to those expected to come
from the Sun, even if the laboratory rate of ion bombardment is,
obviously, by far larger. Thus, if the ion-implantation process is not
the main cause of weathering, one should find a theoretical
explanation of it. \\
\noindent
In summary, we are facing a process characterized by two reddening
timescales: the former, of the order of a few Myr (or even less),
is characterized by significant reddening of several young (and
typically small) objects; the latter, of the order of $10^2-10^3$~Myr,
is characterized by a further reddening, towards saturation of the
effect. While there is no reason to suggest that two different
microphysical effects are at work (why is the faster one not able to
redden after the first few Myr?), the possibility of a complex
process, in which weathering and de-weathering effects are
simultaneously active, seems to deserve  serious scrutiny.\\

\section{Gardening vs. weathering: the model}

The timescale of collisional gardening is not easy to compute, since
it depends on several (partially unknown) parameters. This is also the
basic reason for the forthcoming numerical simulations (see
Sect.~5). Recent estimates, presented in the literature, differ by
several orders of magnitude \citep{wil10,mel09} and it is not easy to
solve the apparent discrepancies. In this Section, rather than present
a new computation, we identify the main parameters of the
problem. Essentially, the basic questions are:

\begin{itemize}
\item{{\bf Q1.} For a given asteroid size, what is the ratio between
  the collisional disruption timescale ($t_{cd}$) and the global
  resurfacing timescale -not automatically entailing a general spectral
  refreshing (see later in the text)- due to an individual impact
  ($t_{\mathrm{gr}}$)?}
\item{{\bf Q2.} Given the size distribution of the projectiles, what
  is the gardening timescale ($t_{\mathrm{ga}}$) due to all
  projectiles?  The result is obtained by integrating between a
  maximum projectile size ($D_{\mathrm{max}}$) and a minimum size
  ($D_{\mathrm{min}}$). When is the integral dominated by the value of
  $D_{\mathrm{min}}$?}
\item{{\bf Q3.} What is the minimum impactor size to be considered?}
\item{{\bf Q4.} If a crater is formed on a reddened region, what is
  the fraction of the ejecta which, refalling onto the asteroid
  surface, gives rise to a -partially or completely- reddened surface?
  This possibility has not been taken into account in previous
  computations. The relevance of the effect depends, obviously, on the
  size of the asteroid, thus introducing differences between small and
  large objects.}
\end{itemize}

\noindent
Let us consider {\bf Q1} first. According to the analysis presented in
\cite{wil10} and \cite{mel09}, the size of the resurfaced region
($d_{\mathrm{res}}$) is roughly proportional to that of the impacting
projectile ($D$), namely $d_{\mathrm{res}} \simeq \alpha D$. Thus the
resurfaced area is $\alpha^2$ times larger than the projectile cross
section. The value of $\alpha$ depends on various parameters, but the
linear relation seems rather reasonable and robust for the typical
conditions of asteroidal impacts: in fact the size of the resurfaced
region is proportional to the size of the crater \citep{mel89}, and
this latter is proportional to that of the projectile \citep[at least
  in the strength regime;][]{mel89}.  Thus we can adopt this as a
basic rule. Consequently, it is possible also to discuss the first
question: a global resurfacing follows from the impact of a projectile
whose size is about $2/\alpha$ times that of the target (the factor
$2$ comes out from the factor $4$ relating the surface of a sphere and
the area of a circle with the same radius).\\
\noindent
The value of $\alpha$ is not easy to estimate. According to
\cite{wil10}, and references therein, the size of the crater
$d_{\mathrm{cra}} \simeq 13 D$, and the size of the resurfaced region
is about $2.3$ times larger than the crater size. The mean size of the
resurfaced region is thus about $30$ times that of the projectile (or,
in terms of area vs.\ the cross-section of the projectile, we have a
factor around $10^3$). However, the size of the crater might be
larger: for instance, the scaling laws suggest a value approximately
twice as large; moreover, the Deep Impact experiment \citep{ric07}
suggests a transient crater, say, more than 100 times larger than the
projectile. On the other hand, to compute the ratio between the
resurfaced area and the crater area is even more difficult: the
estimates might be a factor $\simeq 5$ \citep{wil10}, or $\simeq 25$
\citep{mel09}, which is intermediate between those suggested by 
\cite{gil02}, according to which the ratio between the sizes (to be
squared to convert to areas) is between $2$ and $10$. The above
uncertainties, all together, may affect the value of $\alpha$,
increasing it from a minimum value of about 30 (see above), by even
more than an order of magnitude, consequently strongly decreasing the
gardening timescale.\\
\noindent
Therefore, the minimum projectile/target mass ratio causing a complete
gardening is $\simeq 3\times10^{-5}$ in the conservative case, while going
down to $10^{-8}$ in the extreme opposite; the former value is,
according to current collisional theories \citep{bot05,hol09}, smaller
than that causing a catastrophic disruption but larger than that
causing a complete shattering of the target \citep{wil10}. If, as
usual, the size distribution is a monotonic decreasing function of
size, it entails $t_{\mathrm{gr}} < t_{\mathrm{cd}}$ (maybe
$t_{\mathrm{gr}} \ll t_{\mathrm{cd}}$). Moreover, especially when
considering large targets, the role of gravity has to be taken into
account.\\
\noindent
Note that in order to evaluate the global effects of collisions, it
may also be important to consider the so-called ``global-jolt'',
namely regolith displacement over the entire target surface due to
impact-generated seismic waves. These effects can be estimated
according to \cite{obr06}, by the following equation:

\begin{equation}
  D^{*}(g/g_{\mathrm{t}}) \propto (d_{\mathrm{t}}^{5}/v^{2} )^{1/3} 
\end{equation}
\noindent
where $D^{*}$ is the size of the projectile producing, as a
consequence of the impact, an acceleration $g$ (in units of the
acceleration of gravity $g_{\mathrm{t}}$) to all surface particles,
and is a function of the target diameter $d_{\mathrm{t}}$ and of the
impact speed $v$. Assuming that similar consequences follow from a
similar value of $g/g_{\mathrm{t}}$, the above formula indicates that
the size of a projectile able to produce global-jolt scales as
$d_{\mathrm{t}}^{5/3}$.  Therefore, for the same impact speed, smaller
asteroids are affected by progressively smaller impactors than larger
ones.\\
\noindent
Previous arguments showed that the physics of the impact processes
plays a major role in understanding the collisional disruption rate
and the rate of global resurfacing. Since the outcome of a collision
depends on the physical properties of the target, maybe a single
recipe for all asteroids does not hold. More likely, small rubble-pile
asteroids will have a different response compared to large rubble
piles or monoliths.\\
\noindent
The answer to question {\bf Q2} requires the choice of an asteroid
size-frequency distribution. The problem is not simple and will not be
discussed in detail here. However, we can take a simple power law,
such as the traditional ``Dohnanyi slope'' (Dohnanyi, 1969) obtained
by simplified assumptions regarding a stationary collisional cascade:

\begin{equation}
  dN = A m^{-q}\,dm = A'D^{-q'}\,dD 
\end{equation}
where $A$ and $A'$ are constant quantities and $m$ is the mass. In general 
$q' = 3q -2$;
for the Donhanyi slope $q=11/6$, $q'=7/2$.  If we combine this
assumption with the above-quoted ansatz $d_{\mathrm{res}} = \alpha D$
(and thus $S_{\mathrm{res}} = (\pi/4) \alpha^2 D^2$; $S_{\mathrm{res}}$ is the
resurfaced surface) we obtain the relation:
\begin{equation}
  dS_{\mathrm{res}} = S_{\mathrm{res}}(D) dN \propto D^{-3q+4}\,dD .
\end{equation}

The integral has to be performed between a maximum size
$D_{\mathrm{max}}$ of the order of that causing the collisional
disruption (with a corresponding timescale $t_{\mathrm{coll}}$), or,
better, of the order of the projectile size that causes global
resurfacing $D_{\mathrm{gr}}$, and a minimum $D_{\mathrm{min}}$ to be
defined. With the above assumptions, we obtain something proportional
to $D_{\mathrm{min}}^{-3q+5} - D_{\mathrm{gr}}^{-3q+5}$. If so,
whenever $q > 5/3$ (or, equivalently, whenever the differential size
distribution has an exponent steeper than $-3$), the small impacts
dominate the timescale:

\begin{equation}
  t_{\mathrm{ga}} \simeq t_{\mathrm{gr}} (D_{\mathrm{gr}}/D_{\mathrm{min}})^{3q-5} < t_{\mathrm{coll}} (D_{\mathrm{max}}/D_{\mathrm{min}})^{3q-5}.
\end{equation}

Assuming a Donhanyi slope ($3q-5 = 1/2$), for an asteroid of size a
few km, which should be disrupted by a $\simeq$ km-sized projectile,
the gardening time is of the order of $t_{\mathrm{gr}}/20$ (or less)
if the minimum useful size of the impactor, to cause damage, is
assumed of the order of a few meters \citep[as in][]{wil10}, while
decreasing below $10^{-3} t_{\mathrm{coll}}$ assuming
$D_{\mathrm{min}} \simeq$ 1~mm \citep[as in][]{mel09}. The values can
change also assuming a different size distribution or altering other
parameters of the model.\\
\noindent
Figure~2 compares the analytic distribution with that of MBAs derived
by a collisional evolution model \citep{bot05}.  The plot also shows
the MBA size distribution derived by SUBARU observations
\citep{yos07}, which is valid down to $\sim1$~km. All distributions
are normalized at 5~km. The plot shows that the Dohnanyi slope is a
good approximation of the MBA size-frequency distribution for the
purpose of this work, although, there may be significant local slope
variations.\\
\noindent
The discussion above shows that gardening is usually dominated by
small impacts and that the estimate of the minimum
impactor size causing resurfacing ({\bf Q3}) is crucial for estimating
the gardening timescale.\\
\noindent
The last question of the list ({\bf Q4}), however, may affect the
above conclusions.  In the case of an impact followed by the recapture
of all the ejected fragments, one may imagine that a non-negligible
fraction of the fragments ($\sigma$) will show an already-reddened
surface. This fraction should be particularly large in the limit of a
very small crater. We assume that a newly formed crater excavates
material from pre-existing ejecta layers. This material will be
mixed-up and then spread around to form a new ejecta blanket.  Let
$\beta$ be the volume ratio of reddened particles to non-reddened
particles present in the ejected material ($\beta=0,1$ for
non-reddened and totally reddened particles, respectively).  Thus, the
resulting fraction of reddened particles exposed in the new ejecta
blanket is $\sigma = 0.5\beta$, assuming that half of the particles
should fall showing the reddened surface.\\
\noindent
If we neglect this effect, the maximum attainable spectral slope, due
to the combined effect of weathering (with a timescale of
$t_{\mathrm{sw}}$) and collisions, should be a function of
$t_{\mathrm{sw}}/t_{\mathrm{ga}}$ \citep{wil08,wil10}. A similar
analytic toy model is discussed in \cite{pao09}. Conversely, if we
take it into account, one can imagine a progressive slope saturation,
with a timescale of the order of $t_{\mathrm{ga}}/\sigma$. However, in
the meantime, larger impacts take place, affecting deeper -and
fresher- regions; the resurfacing causes a partial refreshing (the
fraction is related to the fraction of deep, unweathered ejecta) and
the slope does not completely saturate. After a longer time, these
layers also become completely or extensively reddened, and the
saturation is limited only from the consequences of even larger -and
rarer- events. Thus the slope saturation may take an asymptotic
behavior; one can guess that the final timescale to approach the slope
saturation may be of the order of $t_{\mathrm{cd}}$ \citep{pao09}. In
principle one might suggest that the reddest asteroids should be those
whose typical age (or lifetime) is close to the age of the Solar
System. Smaller asteroids should be bluer -for the reasons discussed
above- than intermediate sized bodies; the same holds true for the
larger bodies being farther from the asymptotic saturation.  In the
next Section we compare our ideas with observations.

\section{Comparison with the observations}

The discussion presented in the previous Section is rather
challenging. Some suggestions may be verified only with the aid of a
detailed model of surface evolution, requiring numerical
simulations. In this Section we introduce some observational data that
seem to support, at a semi-quantitative level, our ideas.\\

One of the most interesting points in our model is, certainly, the
different behavior between small and large asteroids. For the former,
gardening is more effective, since the reaccumulation of partially
reddened ejecta is absent or strongly reduced. If the asteroids are a
little bit larger, gardening is less effective, and the surface, for a
given exposure, might be redder. This trend should continue as far as
the asteroid size increases, reaching a plateau when the
reaccumulation becomes massive or almost total. The value of the
corresponding size depends on the physics of collisional processes; it
has certainly to be within the $1-10$s of km size range, since
reaccumulation requires the typical speeds of the ejecta to be smaller
than the escape speed.\\

For larger bodies, as a first approximation, we imagine the spectral
slope to remain close to the plateau value, but, if the arguments
presented at the end of the last section are valid, the complete
``slope saturation'' is reached in a time which is of the order of the
collisional lifetime. Thus, if the lifetime is larger than the
possible age of the body (which cannot exceed the age of the Solar
System), the slope might be slightly under the saturation value. The
transition is for bodies of the order of a few tens of kilometers
\citep[see, for instance,][]{bot05}. In Fig.~\ref{f3} we show the
exposure-corrected slope as a function of size for MBAs.  The
exposure-corrected slope is computed multiplying the observed slope of
a particular object by the ratio between its estimated exposure
(function of the orbital parameters and of the age) and the mean
(among all asteroids) exposure. Thus, the obtained exposure-corrected
slopes are scattered around an average value with no correlation with
the asteroid sizes.  We find a statistically significant increase up
to around 15~km (2-tailed probability of $2.2\times 10^{-11}$),
followed by a significant decrease beyond 15~km (2-tailed probability
of $3.5\times 10^{-9}$). Note that the exposure, used to obtain the
data represented in the figure, assumes constant orbital elements
($a,e$) and an age computed according to the prescriptions of
\cite{mar06a} (applied to all objects regardless their family
membership). However, for family asteroids a different age estimate,
obtained from the analysis of the overall properties of the related
family, can be obtained. We verified that the quality of the trend is
not affected by the definition of the age. With all the uncertainties
in the model, Fig.~\ref{f3} supports our ideas.

\section{Numerical simulations}

As discussed in the previous sections, the basic problem to be solved
for a reliable assessment (or falsification) of the ideas sketched
above, and partially supported by observational evidence, is to
understand what happens when a crater is formed on the surface of an
asteroid. In particular, the critical issues are:

\begin{itemize}

\item{ {\bf I1.} What happens when the projectile impacts onto a
  fractured (regolith, imbricated) surface and the size of the
  projectile is smaller or comparable to that of the surface
  components? (This projectile size is represented by
  $D_{\mathrm{min}}$ in Sect.\ 3.)}
\item{ {\bf I2.} What is the size and velocity distribution of the
  ejecta (and, possibly, the size-velocity relation); in other words,
  given the size of the target, how many -and which- ejecta will fall
  again onto the target, and where (within the crater, close to the
  crater, everywhere on the surface)?}
\item {{\bf I3.} Is there any effect due to the rotational properties
  of the ejecta? In other words, do they, re-falling, show
  approximately the same external surface in about 50\% of the cases
  (consistent with common sense), or is there some subtle reason for a
  different value?}
\end{itemize}

The handling of these problems, with the use of numerical simulations,
might be successful, as indirectly shown by the possibility of
explaining some observed features of Eros in terms of a dynamical
model of crater ejecta \citep{dur09}.\\
\noindent
Direct simulation of regolith dynamics is becoming more feasible with
advances in computer speed and numerical algorithms.  \cite{san11} and
\cite{ric11} have begun development of discrete element methods
tailored for the low-gravity environment of small asteroids.  Key
advances include efficient collision handling, proper accounting of
surface friction, and implementation of weak non-gravitational forces
that may play a critical role in the evolution of surface regolith. \\
\noindent
Many parameters are uncertain, even if critical information and
constraints may come out from the analysis of space missions
\citep[see, for instance,][]{ric07}. However, new interest in sample
return from and possible human exploration of small asteroids is
spurring development of laboratory and computer experiments that may
ultimately shed light on the weathering processes being discussed
here.\\
\noindent
We will devote a forthcoming paper to the implementation of the
required numerical simulations and to the discussion of the results.
Simulations being developed will allow a portion of the granular
surface of an asteroid to be modeled as a collection of discrete,
possibly non-spherical particles in resting contact. Impacts and/or
seismic shaking of the region will be simulated to determine the
extent of ejecta redistribution and overturning as a function of
impactor size, speed, and incidence angle.\\

\section{Conclusions and future work}
\indent

The purpose of the present series of papers is to establish a complete
model of the space weathering processes for S-complex asteroids, in
terms of a balance between the reddening due to weathering and the
refreshment of the surface due to collisional processes, introducing
also the possibility of a reduced refreshment arising from the
reaccumulation of collisional ejecta.  In this paper we have outlined
the general features of the scenario and the most relevant
uncertainties of the theory.  In spite of these uncertainties, we have
suggested a way to overcome the apparent conundrum of the relevant
timescales coming out from the data. The suggested explanation is also
-at least qualitatively- supported by a particular analysis of the
data.  However, the overall complexity of the
impact/cratering/ejection/reaccumulation processes requires a more
detailed analysis in terms of numerical simulations, which we intend
to perform.\\
\noindent
The realistic goal of this future study is a more quantitative
estimate of the involved timescales, allowing a better general model
and a more detailed statistical analysis of the data. However, we
guess that the detailed interactions between external disturbances and
the surface properties of any individual asteroid may lead to a wide
spread of results. Their interpretation, and the related possibility
of deriving some relevant surface properties for a given object,
requires very sophisticated modeling.  In this sense, the verification
or falsification of the suggestions presented in the last Section may
be of some preliminary use.\\
\noindent
Finally, we remark that a reliable model of space weathering of
asteroids is not only interesting {\it per se}, but also represents a
fundamental step to understanding the overall surface evolution of
asteroids, including collisions, thus cratering and erosion, formation
of regolith layers, and so on.

\newpage

\begin{figure}
\vspace{0.5cm}
\includegraphics[width=11cm,angle=-90]{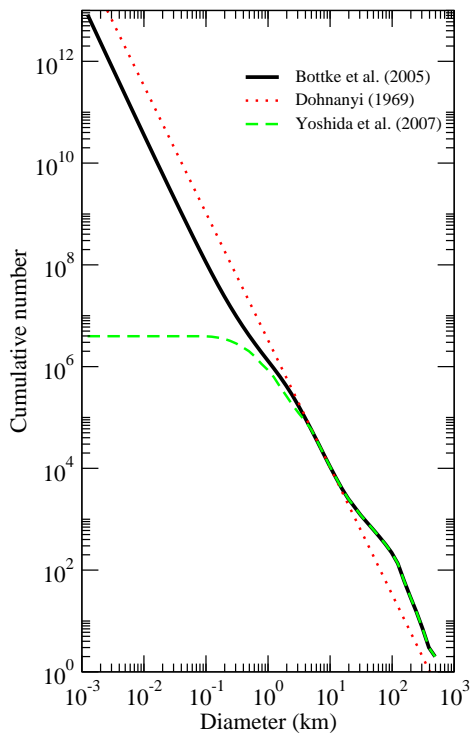}
\caption{Main belt asteroids cumulative size-frequency distributions
  according to Bottke et al.~(2005) model (solid black), to SUBARU
  observations (green, significant only for bodies larger than 1~km,
  Yoshida et al.~2007), and to the analytic Donhanyi relation
  (red). All curves are normalized at 5~km.}
\label{f2}
\end{figure}

\begin{figure}
\vspace{0.5cm}
\includegraphics[width=9cm,angle=-90]{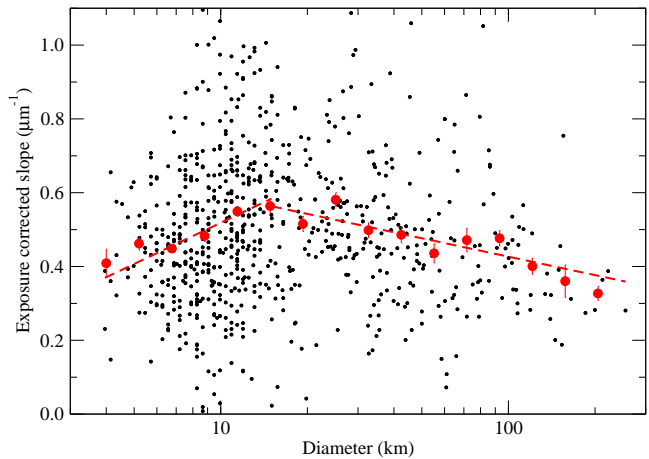}
\caption{The exposure-corrected slope for the S-complex main belt
  asteroids as a function of size. The sample is limited and
  incomplete at the small size end, due to observational biases.  Red
  dots indicate average slope values for binned intervals of asteroid
  sizes.  A best fit has been performed with a linear trend increasing
  up to 20~km and a following decrease.  Both trends are statistically
  significant.  See text for discussion.}
\label{f3}
\end{figure}


\section*{Acknowledgments}

DCR acknowledges support from the National Aeronautics and Space
Administration under Grant No.~NNX08AM39G issued through the Office of
Space Science.  PP has been funded by ASI.  Authors are grateful to
the referee M.J.~Gaffey for useful comments.

\bsp

\label{lastpage}

\end{document}